\documentclass[11pt]{article}
\usepackage{amsmath}
\usepackage[dvips]{graphicx}
\usepackage[dvips]{epsfig}
\usepackage{wrapfig}
\newcommand{\beqann}{\begin{eqnarray*}}  \newcommand{\beqa}{\begin{eqnarray}}
\newcommand{\beqnn }{\begin{displaymath}}\newcommand{\beq }{\begin{equation}}
\newcommand{\btab}{\begin{tabular}}
\newcommand{\etab}{\end{tabular}}

\newcommand{\mc}{multiple comparisons }
\newcommand{\eeqann}{\end{eqnarray*}} \newcommand{\eeqa}{\end{eqnarray}}
\newcommand{\eeqnn}{\end{displaymath}}
\newcommand{\eeq}{\end{equation}}

\newcommand{\Nor}{\mbox{N}}

        \renewcommand{\bibitem}{\vskip
  2pt\par\hangindent\parindent\hskip-\parindent} \renewcommand{\bibitem}{\vskip
  2pt\par\hangindent\parindent\hskip-\parindent}

\author{
  Andrew Gelman\footnote{Department of Statistics and Department of Political Science, Columbia University, New York, {\tt gelman@stat.columbia.edu}, {\tt www.stat.columbia.edu/$\sim$gelman}},
  Jennifer Hill\footnote{Department of Humanities and Social Sciences, NYU Steinhardt, New York University, New York, {\tt jennifer.hill@nyu.edu}}, 
  Masanao Yajima\footnote{Department of Statistics, University of California, Los Angeles, {\tt masanao.yajima@gmail.com}, {\tt http://www.stat.ucla.edu/$\sim$yajima}}}

\title{Why we (usually) don't have to worry about multiple comparisons\footnote{We thank the participants at the NCEE/IES multiple comparisons workshop for helpful comments and the National Science Foundation, National Institutes of Health, and Columbia University Applied Statistics Center for financial support.}}

\date{July 13, 2009}
\begin{document}

\maketitle
\baselineskip=18pt

\begin{abstract}

Applied researchers often find themselves making statistical
inferences in settings that would seem to require multiple comparisons
adjustments.  We challenge the Type I error paradigm that underlies
these corrections.  Moreover we posit that the problem of multiple
comparisons can disappear entirely when viewed from a hierarchical
Bayesian perspective.  We propose building multilevel models in the
settings where multiple comparisons arise.

Multilevel models perform
partial pooling (shifting estimates toward each other), whereas
classical procedures typically keep the centers of intervals
stationary, adjusting for multiple comparisons by making the intervals
wider (or, equivalently, adjusting the $p$-values corresponding to
intervals of fixed width).  Thus, multilevel models address the
multiple comparisons problem and also yield more efficient estimates,
especially in settings with low group-level variation, which is where
multiple comparisons are a particular concern.

%Multilevel estimates make comparisons more conservative, in
%the sense that intervals for comparisons are more likely to include zero; as
%a result, those comparisons that are made with confidence are more likely to
%be valid.

%This paper addresses the dangers of multiple comparisons.  We outline
%how the problem is viewed from a Classical perspective and then
%demonstrate how it can disappear when viewed from a Bayesian
%perspective.  We propose building multilevel models in the settings
%where multiple comparisons arise.  These not only address the multiple
%comparisons problem, they tend to yield more reliable estimates as
%well.

Keywords:  Bayesian inference, hierarchical modeling, multiple comparisons, Type S error, statistical significance

\end{abstract}

\section{Introduction}
Researchers from nearly every social and physical science discipline
have found themselves in the position of simultaneously evaluating
many questions, testing many hypothesis, or comparing many point
estimates.  In program evaluation this arises, for instance,
when comparing the impact of several different policy interventions,
comparing the status of social indicators (test scores, poverty rates,
teen pregnancy rates) across multiple schools, states, or countries,
examining whether treatment effects vary meaningfully across different
subgroups of the population, or examining the impact of a program on
many different outcomes.
% this list doesn't cover all our scenarios but I think that's ok.

The main multiple comparisons problem is that the probability a
researcher wrongly concludes that there is at least one statistically
significant effect across a set of tests, even when in fact there is
nothing going on, increases with each additional test.  This can be a
serious concern in classical inference, and many strategies have been
proposed to address the issue (see Hsu, 1996, and Westfall and Young,
1993, for reviews).  A related multiple comparisons concern is that, in
a setting where nonzero true effects do exist for some of the
phenomena tested, a researcher applying multiple tests may identify
additional statistically significant effects that are not in fact
real.

%Nonetheless, we usually don't get upset about multiple comparisons in our applied work.  As described above, multiple comparisons seem like a major worry,
%but from another perspective, they don't matter at all:
%\begin{itemize}
%  \item They are typically not a concern when we study phenomena with non-zero effects
%  \item They are typically not a concern if we study comparisons with non-zero differences
%  \item We don't mind being wrong 5\% of the time. (THIS ONLY FITS WITH CERTAIN TYPES OF MC PROBLEMS)
%\end{itemize}
%This might sound glib but we really believe it, and Bayesians put it into practice all the time, as can be seen, for example, from the dozens of examples in our books (Gelman et al., 2003, Gelman and Hill, 2007).  need better evidence than US putting it into practice

Our approach, as described in this paper, has two key differences from
the classical perspective.  First, we are typically not terribly
concerned with Type 1 error because we rarely believe that it is
possible for the null hypothesis to be strictly true.  Second, we
believe that the problem is not multiple testing but rather insufficient modeling of
the relationship between the corresponding parameters of
the model.  Once we work within a Bayesian multilevel modeling
framework and model these phenomena appropriately, we are actually
able to get more reliable point estimates.  A multilevel model shifts
point estimates and their corresponding intervals toward each other
(by a process often referred to as ``shrinkage'' or ``partial
pooling''), whereas classical procedures typically keep the point
estimates stationary, adjusting for multiple comparisons by making the
intervals wider (or, equivalently, adjusting the $p$-values
corresponding to intervals of fixed width).  In this way, multilevel
estimates make comparisons appropriately more conservative, in the
sense that intervals for comparisons are more likely to include zero.
As a result we can say with confidence that those comparisons made
with multilevel estimates are more likely to be valid.  At the same
time this ``adjustment'' doesn't sap our power to detect true
differences as many traditional methods do.

Rather than correcting for the problems that can arise when examining
many comparisons (performing many significance tests), when we work
within the Bayesian paradigm all of the relevant research questions
can be represented as parameters in one coherent multilevel model.
Simply put, rather than correcting for a perceived problem, we just
build the multiplicity into the model from the start.  This puts more of a burden
on the model, and a key goal of this paper is to demonstrate the
effectiveness of our procedure in realistic examples.

Sections 2 and 3 present the multiple comparisons problem from the
classical and Bayesian perspectives, respectively.  Both are described
within the context of a common illustrative example and then potential
solutions are outlined.  In Section 4, we bolster our argument against traditional
\mc corrections through a series of small
examples that illustrate several of the scenarios described above.
Section 5 concludes.
%For each we describe the potential problem, the shortcomings of a traditional
%multiple comparisons ``fix'', and the multilevel model solution.

\section{Multiple comparisons problem from a classical perspective}

\subsection{Illustrative example}\label{illustrative}
In this section we walk through a relatively simple example using data
from a real study to illustrate the issues involved in performing
multiple comparisons from classical and 
multilevel perspectives.  We use data from the Infant Health
and Development Program, an intervention that targeted premature and low-birth-weight infants and provided them with
services such as home visits and intensive high quality child care.
The program was evaluated using an experiment in which randomization
took place within site and birth weight group.
% (low is less than or
%equal to 2000 grams, high is between 2000 and 2500) CITATIONS.
The experimental design was actually slightly more complicated (as
described in IHDP, 1990); we simplify here for
expository purposes.  In fact, for this first illustration we will
assume that it was a simple randomized block experiment with eight
sites as blocks.

In this context, we're not just interested in the overall treatment
effect.  Given that the composition of participating children was
quite different across sites and that program implementation varied
across sites as well, we would like to know for \emph{each site
individually} whether or not a statistically significant effect was
present.  However, we may be concerned that, in the process of
conducting eight different significance tests, we are misperceiving our
overall risk of making a false claim.  This overall risk of error
(formally, the probability that we have any rejections when the null
hypothesis in fact holds) is sometimes referred to as the familywise
error rate (Tukey, 1953).  A similar problem arises if we are
interested in comparing whether there are significant differences in
treatment effects across sites.

\subsection{Classical perspective}

A classical model fit to these data might look like:
\begin{eqnarray*}
y_i &=& \sum_{j=1}^8 (\gamma_j S^j_i + \delta_j S^j_i P_i) + \epsilon_i, \\
\epsilon_i &\sim& \Nor(0,\sigma^2),
\end{eqnarray*}
%\alpha + \tau P_i + \sum_{j=2}^8 (\gamma_j S^j_i + \delta_j S^j_i P_i)
%In this commonly-used specification $\tau$ represents the treatment
%effect for the first site, and $\delta_j$ ($j=2,\ldots,8$) represents
%the difference between the treatment effect in the $j^{\text{th}}$ site
%and the first site
where $y_i$ denotes student $i$'s test score, $S^j_i$ in an indicator
for living in site $j$, and $P_i$ is an indicator for program status.
Although this may not be the most common way to specify this model, it
is useful here because $\delta_j$ represents the treatment effect in
the $j^{\text{th}}$ site and $\gamma_j$ represents the average test
score for the untreated in each site.\footnote{The actual analysis also included birthweight
group as a predictor in this model, but we ignore this in
this description for simplicity of exposition.}  This allows us to
directly test the significance of each site effect.

For any given test of a null hypothesis, say $H^j_0: \delta_j = 0$,
versus an alternative, say $H^j_A: \delta_j \ne 0$,
% \mbox{H}_{\mbox{A}}  looked kind of crappy
there is a 5\% chance of incorrectly rejecting $H^j_0$ when in fact it
is true.  Of course if we test two independent hypotheses at the same
significance level ($\alpha$ = 0.05) then the probability that at least
one of these tests yields an erroneous rejection raises to
$1-\Pr(\text{neither test yields an erroneous rejection of the null})
= 1-0.95*0.95 = 0.098 \approx 0.10$.  Following the same logic, if we
performed (independent) tests for all eight sites at a 0.05 significance
level there would be a 34\% chance that at least one of these would
reject in error.

\subsection{Bonferroni correction}
\begin{figure}
% \vspace{-0.5in}
%  \hspace{0.2in}
  \centerline{\psfig{figure=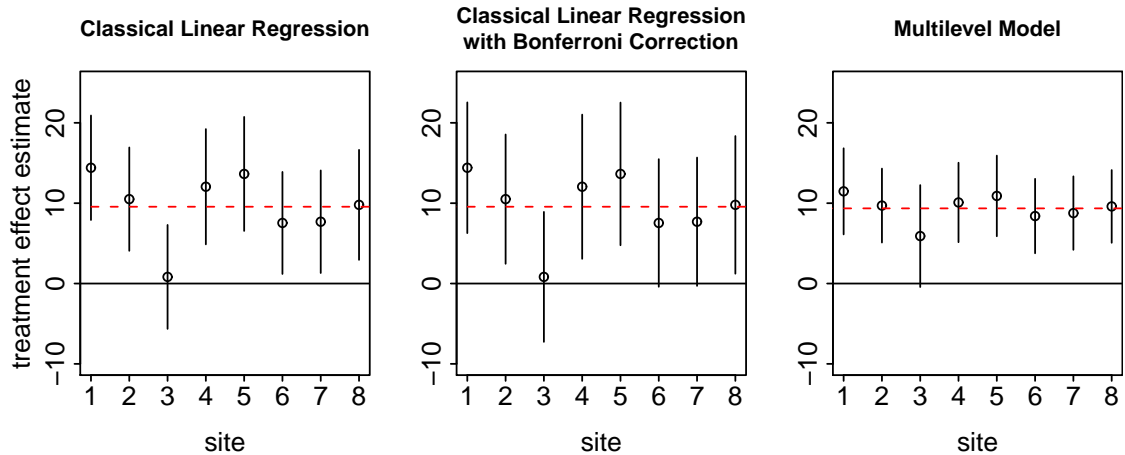,height=2.5in}}
%  \centerline{\psfig{figure=plot1.eps,height=2.5in}}
  \caption{Treatment effect point estimates and 95\% intervals across the eight  Infant Health
and Development Program sites.  The left panel display classical estimates from a linear regression.  The middle panel displays the same point estimates as in the left panel but with confidence intervals adjusted to account for a Bonferroni correction.  The right panel displays posterior means and 95\% intervals for each of the eight site-specific treatment effects from a fitted multilevel model.
}
  \label{ihdp}
\end{figure}

One of the most basic and historically most popular fixes to this
problem is the Bonferroni correction.  The Bonferroni correction
adjusts the $p$-value at which a test is evaluated for significance
based on the total number of tests being performed.  Specifically, the
working $p$-value is calculated as the original $p$-value divided by the
number of tests being performed.  Implicitly, it assumes that these
test statistics are independent.  So in our current example an overall
desired significance level of 0.05 would translate into individual
tests each using a $p$-value threshold of $0.05/8 = 0.0062$.  These
thresholds could also be used to create wider confidence intervals for
each point estimate as displayed in Figure \ref{ihdp}, which plots the point
estimates from the model above along with both uncorrected and
Bonferroni-corrected uncertainty intervals corresponding to a nominal
0.05 significance level.  While the standard intervals reject the null
hypothesis of no effect of the intervention for 7 of the 8 sites, the
multiple-comparisons-adjusted intervals reject the null hypothesis for
only 5 sites.

The Bonferroni correction directly targets the Type 1 error problem, but
it does so at the expense of Type 2 error.  By changing the $p$-value
needed to reject the null (or equivalently widening the uncertainty
intervals) the number of claims of rejected null hypotheses will
indeed decrease on average.  While this reduces the number of false
rejections, it also increases the number of instances that the
null is not rejected when in fact it should have been.  Thus, the
Bonferroni correction can severely reduce our power to detect an
important effect.

\subsection{Other classical corrections}
Motivated by the shortcomings of the Bonferroni correction, 
researchers have proposed more sophisticated procedures.  The goal of
these methods typically is to reduce the familywise error rate (again,
the probability of having at least one false positive) without unduly
sacrificing power.  A natural way to achieve this is by taking account
of the dependence across tests.  A variety of such corrections exist
that rely upon bootstrapping methods or permutation tests (see, for
example, Westfall and Young, 1993).

%there are some circumstances where multiple comparisons aren't really a
%problem after all (even though XXX).

A more recent class of approaches to this problem focuses not on
reducing the familywise error rate but instead on controlling the
expected proportion of false positives, the ``false discovery
rate'' or FDR (Benjamini and Hochberg, 1995).  The rationale
is that the researcher should be
more worried about a situation in which many tests show up as
statistically significant and an unknown proportion of these are
erroneous than a situation in which all but a few tests show up as
insignificant.  Controlling for the false discovery rate rather than
the familywise error rate leads to a less conservative testing
procedure with respect to Type 1 error but is more powerful in terms
of detecting effects that are real.  These tests sit squarely within
the Type 1 paradigm.  As with procedures to control for the familywise
error rate, the initial versions assumed independence across tests;
however, procedures to control the FDR have also been extended to
account for certain types of dependency across tests (Benjamini and
Yekutieli, 2001).

Methods that control for the FDR may make particular sense in fields
like genetics where one would expect to see a number of real effects
amidst a vast quantity of zero effects such as when examining the
effect of a treatment on differential gene expression (Grant et al.,
2005).  They may be less useful in social science applications when
we are less likely to be testing thousands of hypotheses at a time and
when there are less likely to be effects that are truly zero (or at
least the distinction between zero and not-zero may be more blurry).
For our IHDP example, using a standard procedure (the Simes procedure, see Benjamini and 
Hochberg, 1995) to control the FDR at a 0.05 level, the tests with the 6
smallest $p$-values would reject.

\subsection{Informal calibration}
Still others might argue that in many situations there is no
need to formally control for these error rates.  Many of us are
already used to informally performing appropriate calibrations.  
For instance, consider a researcher who presents a table of mean 
differences in pre-treatment variables across an experimental and 
control group in which there is one statistically significant difference.
The classical perspective would have us worry if we performed
20 tests at a 0.05 significance level that there is a 64\% chance that
at least one of these will yield a statistically significant result
inappropriately.  Thus, for instance, a Bonferroni correction could be 
performed to change the critical value to reflect a $p$-value of $0.05/20 = 0.0025$.   
However, it is probably more common (and at least equally helpful)
for the researcher to simply note that given 20 such tests we would
\emph{expect} to see at least one such deviation given a 0.05
significance level.  Alternatively, research organizations sometimes recommending pre-specifying the tests that will be performed in an offically-sanctioned analysis.  In either case, the goal is to manage expectations in a way similar to a mutiple comparisons correction but perhaps less strictly tied to familywise error rates.
%%THIS SAME AS ROTHMAN ARGUMENT?  check for next round

\section{A different perspective on multiple comparisons}
Classical methods typically start with the assumption that the null
hypothesis is true---an unhelpful starting point, as we discuss below.
Moreover, we argue that they classical procedures fail by insufficiently modeling the ensemble of parameters corresponding
to the tests of interest (see Louis, 1984).  We cannot hope to have the goal of proposing an optimal statistical method for all
circumstances.  Rather we present an entirely different perspective on
the issue and its implications and argue that, when viewed from a
Bayesian perspective, many of these problems simply disappear.

\subsection{Abandoning the Type 1 error paradigm}
The classical perspective worries primarily about Type 1 errors,
and we argue that these should not be the focus of our concern.  
Suppose we've established the following two hypotheses
regarding our site-specific treatment effects $\tau_j$ for
$j=1,\ldots,J$: $H^j_0: \tau_j=0$, and $H^j_A: \tau_j \ne 0$.  A
primary concern from the classical multiple comparisons perspective is
that we might erroneously accept $H^j_A$ when, in fact, $H^j_0$ is
true (Type 1 error).  But do we ever believe that $\tau_j$ exactly
equals zero?  What is the practical importance of such a test?  Similarly,
a Type 1 error occurs when we mistakenly accept the $H^j_0$ that
$\tau_j = \tau_k$ when in fact the $H^j_A$ that $\tau_j \ne \tau_k$ is
true.  Again, under what circumstances do we truly believe there are
absolutely no differences between groups?  There may be no
\emph{practical} differences but this is a distinct point which we
shall discuss shortly.  Moreover, if true effects are zero, we don't
want anything close to a 5\% chance of finding statistically
significant results.

A more serious concern might be that we claim $\tau_j>0$ when in fact
$\tau_j<0$, finding a positive effect when in fact the effect is
detrimental.  A similar phenomenon occurs if we claim that
$\tau_j>\tau_k$ when in fact $\tau_j<\tau_k$, for instance claiming
that a treatment effect is larger in Miami than in New York when in
fact the reverse is true.  These are both examples of what is referred
to as ``Type S'' (sign) errors (Gelman and Tuerlinckx, 2000).

%We are mostly interested in examples where differences are
%unquestionably real, and the key concern is Type S errors: for
%example, saying that School A is better than School B, when School B
%is really better than School A.

\begin{figure}
%\begin{wrapfigure}{1}{.66\textwidth}
 \vspace{-0.5in}
%  \hspace{0.2in}
  \centerline{\includegraphics[width=.6\textwidth]{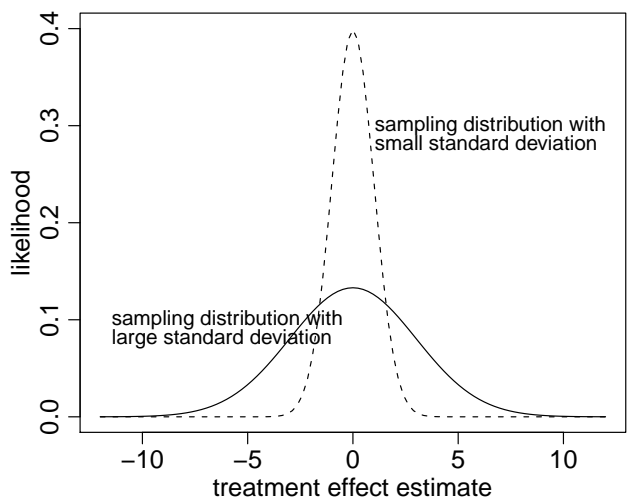}}
%  \centerline{\psfig{figure=two.dists.eps,height=3in}}
\vspace{-.3in}
  \caption{Two error distributions with differing levels of uncertainty in a situation when there is no effect.  The estimator with the sampling distribution with greater uncertainty (standard deviation equal to 3) has a greater probability of yielding a larger estimate than the estimator with the sampling distribution with smaller uncertainty (standard deviation equal to 1).  The researcher is more likely to commit a Type M (magnitude) error when the standard deviation is large.
}
  \label{fig.two.dists}
%\end{wrapfigure}
\end{figure}

However in policy analysis, there is also a fair bit of concern about
examples where the differences might actually be very close to zero:
for example, comparing different educational interventions, none of
which might be very effective.  Here we would want to be thinking
about ``Type M'' (magnitude) errors: saying that a treatment effect is
near zero when it is actually large, or saying that it's large when
it's near zero (Gelman and Tuerlinckx, 2000).  In that setting,
underpowered studies present a real problem because type M errors 
are more likely when uncertainty is high.
%may systematically occur even the subset of estimates that happen to be 
%statistically significant (Gelman and Weakliem, 2009).
For instance it is not uncommon in an underpowered study for a researcher to
state that although his estimate is not statistically significantly
different from 0, that could simply be a function of the overly large
standard error.  Ironically, however, large estimates are actually a
\emph{byproduct} of large standard errors.

This property is illustrated in Figure \ref{fig.two.dists}.  This plot
displays two sampling distribution in a situation in which the true
effect is zero (or very close to 0).  It's clear from this plot that
the estimator with the sampling distribution with greater uncertainty
(standard deviation equal to 3) is much more likely to produce effect
estimates that are larger in magnitude than effect estimates resulting
from an estimator with relatively less uncertainty (standard deviation
equal to 1).  Thus, for instance, when we switch from examining main
effects to subgroup effects, for example, we automatically increase
our probability of seeing large estimates and tricking ourselves into
thinking that something is going on.  Bayesian modeling helps here
too, as we shall see below.

%again, we don't think multiple
%#comparisons adjustments are necessary, since a multilevel model should
%comfortably pool all estimates to close to zero when sample sizes are
%too small to learn anything useful.

%The decision regarding whether the magnitude of an effect, if deemed
%positive, is large enough to merit \emph{practical significance} is
%another key (and controversial) issue. DEF OF TYPE-M ERROR?

\subsection{Multilevel modeling in a Bayesian framework}
More strongly, we claim that when viewed within a Bayesian framework,
many of these problems disappear, or in the case of Type S and Type M errors,
are at least substantially ameliorated.
%We'll illustrate the implications of fitting a Bayesian multilevel model with
We illustrate with a relatively simple multilevel model for a setting in which
individuals in a common site experience the same
effect on test scores, as in
\begin{equation*}
y_i \sim \Nor(\gamma_{j[i]} + \delta_{j[i]}P_i,\sigma^2_y),
\end{equation*}
Here $\delta_{j[i]}$ is the parameter for the treatment effect corresponding
to person $i$'s site (indexed by $j$), and it is assigned its own distribution, for example
\begin{equation*}
\delta_j \sim \Nor(\mu,\sigma^2_{\delta}).
\end{equation*}
We have also allowed the intercept, $\gamma$, to vary across
sites in a similar manner.  It does not seem a strong assumption to
think of these as realizations from a common distribution and this
addition should strengthen our model.  Additionally, our Bayesian analysis
requires us to specify prior distributions for the parameters
$\mu$, $\sigma_y$, and $\sigma_\delta^2$.  However (particularly for this
kind of simple model) it is not difficult to choose weakly informative priors (Gelman, 2006).
Finally, we could (and should, in a real analysis) easily include other predictors to the model to increase
our predictive power---most notably, the group-level intercept $\gamma_j$ can be a predictor for the group-level treatment effect $\delta_j$---but have refrained from adding predictors in this example, so we can focus on primary issues.

%However, we don't think that there is no relationship between these treatment
%effects, therefore we posit a model for the treatment effects as

\paragraph{Partial pooling.}  Multilevel modeling can be thought of as a compromise
between two extremes.  One extreme, complete pooling, would assume the
treatment effects are the same across all sites, that is,
$\delta_j=\delta$, for all $j$.  The other extreme, no pooling, would
estimate treatment effects separately for each site.  The compromise
found in the multilevel model is often referred to as \emph{partial
pooling}.  Figure \ref{ihdp} graphically illustrates this compromise
with a plot of the multilevel intervals next to the classical
estimates and intervals (with and without Bonferroni corrections).
The horizontal dashed line in each plot displays the complete pooling
estimate.  We also display a horizontal solid line at zero to quickly
show which estimates would be considered to be statistically
significant.  This process leads to point estimates that are closer to
each other (and to the ``main effect'' across all sites) than the
classical analysis.  Rather than inflating our uncertainty estimates,
which doesn't really reflect the information we have regarding the
effect of the program, we have shifted the point estimates toward each
other in ways that reflect the information we have.  (More generally,
if the model has group-level predictors, the inferences will be
partially pooled toward the fitted group-level regression surface
rather than to a common mean.)

\paragraph{The intuition.}
Why does partial pooling make sense at an intuitive level?  Let's
start from the basics.  The only reason we have to worry about
multiple comparisons issues is because we have uncertainty about our
estimates.  If we knew the true (population-average) treatment effect within each site, we
wouldn't be making any probabilistic statements to begin with---we
would just know the true sign and true magnitude of each (and
certainly then whether or not each was really different from 0 or
from each other).  Classical inference in essence uses only the
information in each site to get the treatment effect estimate in
that site and the corresponding standard error.

A multilevel model, however, recognizes that this site-specific
estimate is actually ignoring some important information---the
information provided by the other sites.  While still allowing for
heterogeneity across sites, the multilevel model also recognizes that
since all the sites are measuring the same phenomenon it doesn't make
sense to completely ignore what has been found in the other sites.
Therefore each site-specific estimate gets ``shrunk'' or pulled
towards the overall estimate (or, in a more general setting, toward a
group-level regression fit).  The greater the uncertainty in a site,
the more it will get pulled towards the overall estimate.  The less
the uncertainty in a site, the more we trust that individual estimate
and the less it gets shrunk.  

\begin{figure}
%  \centerline{\psfig{figure=shrinking.ps,height=3in}}
%  \centerline{\includegraphics[height=3in,angle=270]{iqsb36.comp.full.eps}}
  \centerline{\includegraphics[height=2.5in]{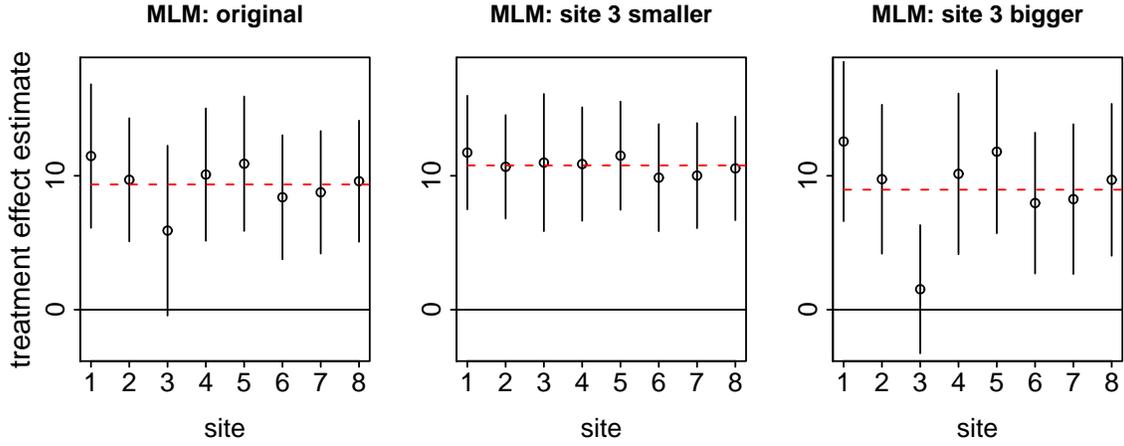}}
  \caption{Comparison of results from multilevel model using different
versions of the data.  The left panel displays results from the original
data.  The center panel displays results from a model fit to data where
Site 3 has been reduced to a random sample of 30 from the original 138
observations in that site.  The right panel displays results from a model
fit to data in which Site 3 observations were bootstrapped to create a
sample of size 300.\label{ihdp.altered}}
\end{figure}

To illustrate this point we ran our multilevel model on slightly 
altered versions of the dataset.  In the first altered version we 
decreased the sample size in Site 3 from 138 to a random sample of
30; results are displayed in the center panel of Figure 
\ref{ihdp.altered}.  In the second altered version we increased the 
sample size in Site 3 to 300 by bootstrapping the original observations 
in that site; results are displayed in the right panel of 
Figure \ref{ihdp.altered}.  The left-most panel displays the original 
multilevel model results.  The key observation is that the shrinkage 
of the Site 3 treatment effect estimate changes drastically across 
these scenarios because the uncertainty of the estimate relative to 
that of the grand mean also changes drastically across these scenarios.  
Note, however, that the overall uncertainty increases in the right-most 
plot even though the sample size in Site 3 increases.  That is because we
increased the sample size while keeping the point estimate the same.
This leads to greater certainty about the level of treatment effect 
heterogeneity across sites, and thus greater uncertainty about the
overall mean.

%For instance in Figure \ref{ihdp}, Site
%2 has the smallest site-specific standard error for its treatment
%effect estimate and Site 3 has the largest.  That is why in the
%display of results from the multilevel model, the Site 3 point
%estimate (posterior mean) is shrunk dramatically towards the common
%treatment effect estimate whereas the Site 2 point estimate (posterior
%mean) barely moves at all.
%% THIS EXAMPLE NOT COMPELLING BECAUSE SITE 2 IS ALREADY SO CLOSE TO
%% THE GRAND MEAN THERE IS NOWHERE FOR IT TO GET SHRUNK TO...
%% PERHAPS RERUN AFTER RANDOMLY KILLING HALF THE OBSERVATIONS IN A SITE
%% TO SEE THE DIFFERENCE IN SHRINKAGE

%It has been recognized since the
%pioneering work of James and Stein (1960) and Efron and Morris (1975)
%that partial pooling typically leads to estimates with better
%properties (for instance lower mean squared error) than those produced
%by traditional estimators.

\paragraph{The algebra.}

Partial pooling tends to reduce the number of statistically significant comparisons.
To see this algebraically, consider the estimate for the treatment effect in a single group in a simple normal-normal hierarchical model:
\begin{equation*}
\mbox{posterior E}(\theta_j) = \left(\frac{1}{\sigma^2_{\theta}}\,\mu + \frac{1}{\sigma^2_{\bar{y}}}\,\bar{y}\right)\left/\left(\frac{1}{\sigma^2_{\theta}} + \frac{1}{\sigma^2_{\bar{y}}}\right).\right.
\end{equation*}
The corresponding uncertainty for this estimate is
\begin{equation*}
\mbox{posterior sd}(\theta_j) = 1\left/\sqrt
{\frac{1}{\sigma^2_{\theta}} + \frac{1}{\sigma^2_{\bar{y}}}}\right.
\end{equation*}

The smaller the prior variance $\sigma^2_{\theta}$, the more the posterior estimates for
different groups are pooled toward a common mean value.  At the same
time, their posterior variances are shrunk toward zero, but much more
slowly.

As a result, the $z$-score for any comparison---the
difference in posterior means, divided by the posterior standard
deviation for that difference---decreases, and statistically
significant Bayesian comparisons become less likely.  Algebraically:

\begin{eqnarray*}
\mbox{posterior E}(\theta_j-\theta_k)&=&\frac{\sigma^2_{\theta}}{\sigma^2_{\bar{y}}+\sigma^2_{\theta}}(\bar{y}_j-\bar{y}_k)\\
\mbox{posterior sd}(\theta_j-\theta_k)&=&\sqrt{2}\sigma_{\bar{y}}\sigma_{\theta}\left/\sqrt{\sigma^2_{\bar{y}}+\sigma^2_{\theta}}\right. \\
\mbox{posterior $z$-score of }\theta_j-\theta_k: &&
\frac{(\bar{y}_j-\bar{y}_k)}{\sqrt{2}\sigma_{\bar{y}}}\cdot
\frac{1}{\sqrt{1+\sigma^2_{\bar{y}}/\sigma^2_{\theta}}}
\end{eqnarray*}

\begin{figure}
%  \centerline{\psfig{figure=shrinking.ps,height=3in}}
  \centerline{\includegraphics[height=4in,angle=270]{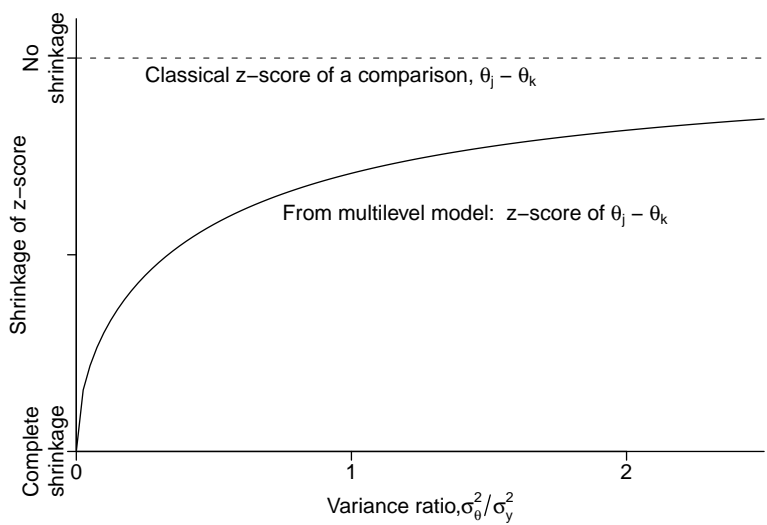}}
  \caption{Shrinkage of the $z$-score for a comparison, $\theta_j-\theta_k$, as estimated using multilevel modeling as compared to classical inference.  When the variance ratio is small---that is, when the groups are similar to each other--Plots of two sampling distributions with differing levels of uncertainty.  When the group-level variance is small, there is a lot of shrinkage.}\label{shrinking}
\end{figure}

The first factor in this last expression is the $z$-score for the
classical unpooled estimates; the second factor is the correction from
partial pooling, a correction that is always less than 1 (that is, it
reduces the $z$-score) and approaches zero as the group-level
variance $\sigma^2_{\theta}$ approaches zero; see Figure \ref{shrinking}.

The actual adjustment
to $z$-scores and significance levels is slightly more complicated
because the variance parameters and the group-level mean are estimated
from data, but the basic pattern holds, which is that posterior means
are pulled together faster than posterior standard deviations
decrease.

Greenland and Robins (1991) make a similar argument about the advantages of 
partial pooling, going so far as to frame the multiple comparisons problem as 
an ``opportunity to improve our estimates through judicious use of any
prior information (in the form of model assumptions) about the
ensemble of parameters being estimated. Unlike conventional multiple
comparisons, EB [empirical Bayes] and Bayes approaches will alter and can improve point
estimates and can provide more powerful tests and more precise
(narrower) interval estimators.''

\paragraph{Model fitting.}
One barrier to more widespread use of multilevel models is that
researchers aren't always sure how to fit such models.  We often
recommend fitting multilevel models in a fully Bayesian way using a
software package such as Bugs (as described in detail in Gelman and
Hill, 2007).  However many simple models can be fit quite well using
packages that have been built in (or can be easily installed into)
existing software packages.  For instance the model above can be fit
easily in {\tt R}, as

\begin{verbatim}
ihdp.fit <- lmer (y ~ treatment + (1 + treatment | group))
\end{verbatim}
Further functions exist in the {\tt arm} package in R to help the user sample from the posterior
distribution for each site-specific treatment effect (or any other
parameter from the model or functions thereof; see Gelman and Hill,
2007).  Similar options for fitting the model are available in Stata
and SAS as well (see Appendix C of Gelman and Hill, 2007). 

If we want to make comparisons across two sites, say site 1
and site 3, we don't need
to refit the model using different contrasts or perform any algebraic manipulations.  All can be done using posterior simulations.

\section{Examples}
We explore our ideas on multilevel models and multiple comparisons
through a series of
examples that illustrate different scenarios in which multiple
comparisons might arise as an issue.
%Since FDR procedures are most appropriate for very large numbers of tests or comparisons, we will only use an FDR procedure as our classical alternative to our preferred analysis in one of these examples.  ** I commented this out as it seems unnecessarily apologetic.  After all, the FDR people would probably say their methods are always appropriate!

\subsection{Comparing average test scores across all U.S. states}

\begin{figure}
 \vspace{-0.5in}
  \hspace{0.2in}
  \centerline{\includegraphics[scale=.65]{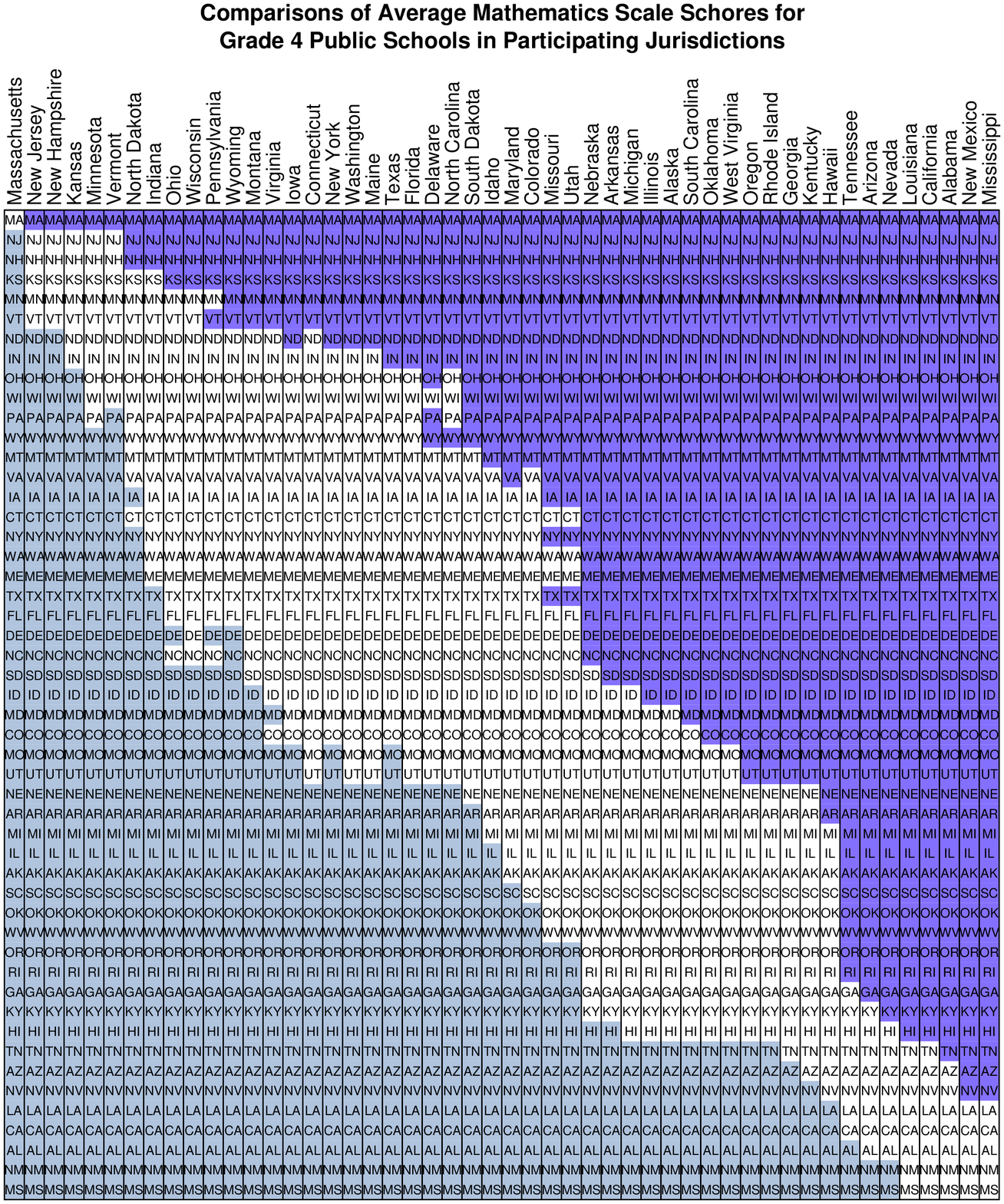}}
%  \centerline{\psfig{figure=naepRecreated.2007.pdf,height=7in}}
  \caption{Graph that mimics one produced for the National Center for Education Statistics (1997) report comparing average math test scores of students in different states.  Shaded comparisons represent differences that are statistically significant after a FDR multiple comparisons correction.  We argue that these multiple-comparisons adjustments make no sense, as they are based on the irrelevant model in which true population differences are exactly zero.}
  \label{naep}
\end{figure}

This next example illustrates how these issues play out in a situation
in which all pairwise comparisons across groups are potentially of
interest.  Figure \ref{naep} shows a graph that mimics one produced
for a National Center for Education Statistics (NCES; 1997) report
that ordered all states based on average scores on the National
Assessment of Educational Progress (NAEP) fourth-grade mathematics
test.  Our version makes use of 2007 fourth-grade mathematics scores
and performs the false discovery rate (FDR) correction currently used by NCES for this
sort of problem (many comparisons).  In the graph, statistically significant
comparisons have been shaded.  In theory, this plot allows us to
answer questions such as, Does North Carolina have higher average
test scores than neighboring South Carolina? This information could
be displayed better (Wainer, Hambleton, and Meara, 1999, Almond et
al., 2000), and maybe should not be displayed at all (Wainer, 1986),
but here our concern is with the formulation as a multiple comparisons
problem.

\paragraph{Concerns with the classical multiple comparisons display.}
Here is a situation in which most classical multiple comparisons
adjustments, such as the FDR adjustment that was used, will not
be appropriate because we know ahead of time that the null hypothesis
(zero average differences between states) is false, so there is no
particular reason to worry about the Type 1 error rate.  Therefore, any
motivation for multiple comparisons then rests either on (a) wanting
more than 95\% of the 95\% intervals to contain the true values, or
(b) wanting a lower Type S error rate, in other words, minimizing the chance
of, for instance, stating that New Jersey has higher average test
scores than Pennsylvania when, in fact, the opposite is the case.

With regard to 95\% intervals, we can do better using multilevel
modeling, either on the raw state averages from any given year or,
even better, expanding the model to include state-level predictors and
test scores from other years.  If Type S error rates are a concern,
then, again, a multilevel model will more directly summarize the
information available.

The objection may be raised that although we know that
the true differences cannot be {\em exactly} zero, what about the
null hypothesis that they are {\em nearly zero}?  Our reply is that
this weaker version of classical hypothesis testing doesn't work here
either.  One way to see this is that the data used to create Figure \ref{naep}
clearly reject either of these null hypotheses.  But classical multiple
comparisons procedures just plug along ignoring this information,
widening the confidence intervals as new comparisons are added.

\paragraph{Multilevel model and corresponding display of comparisons.}

\begin{figure}
%  \vspace{-.8in}
%    \centerline{\includegraphics[scale=0.9]{schools.sim.eps}}
  \centerline{\psfig{figure=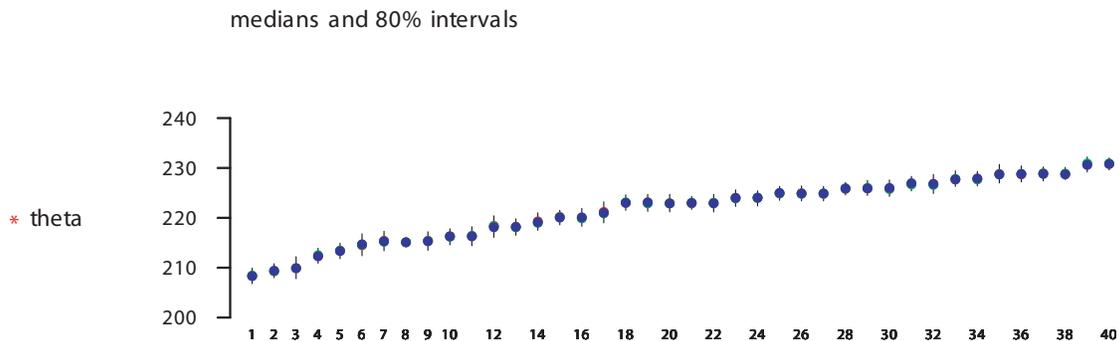,height=2.5in}}
  \caption{Summary of inferences for average math scores by state (see Figure \ref{naep}) based on a fitted multilevel model.  States have been ordered by increasing average test score in the raw data.  This graph could be improved in many ways; actually, though, it portrays the comparisons fairly clearly while being nothing more than a piece of the default graphical output obtained by fitting the model in R and Bugs.\label{naepMultilevel}}
\end{figure}

As an alternative, we fit a multilevel model: $y_{j}\sim\Nor(\alpha_j,
\sigma^2_j)$, where $j=1,\ldots,J$ are the different states, and
$y_{j}$ is the average fourth grade mathematics score for the students
who took the test in state $j$.\footnote{We recognize that this model
could be improved, most naturally by embedding data from multiple
years in a time series structure.  The ability to include additional
information in a reliable way is indeed a key advantage of multilevel
models; however, here we chose a simple model because it uses no more
information than was used in the published tables.}  The parameters
$\alpha_j$ represent the true mean in each state---that is, the
population average that would be obtained if all the students in the
state were to take the test.  We model these population averages with
a normal distribution:
$\alpha_j\sim\Nor(\mu_{\alpha},\sigma^2_{\alpha})$.  Finally, we
assign noninformative uniform prior distributions to the hyperparameters
$\mu_{\alpha},\sigma_{\alpha}, \sigma_j$.  We display the resulting
estimated state-level parameters in Figure \ref{naepMultilevel}.

\begin{figure}
  \vspace{-.8in}
    \centerline{\includegraphics[scale=0.65]{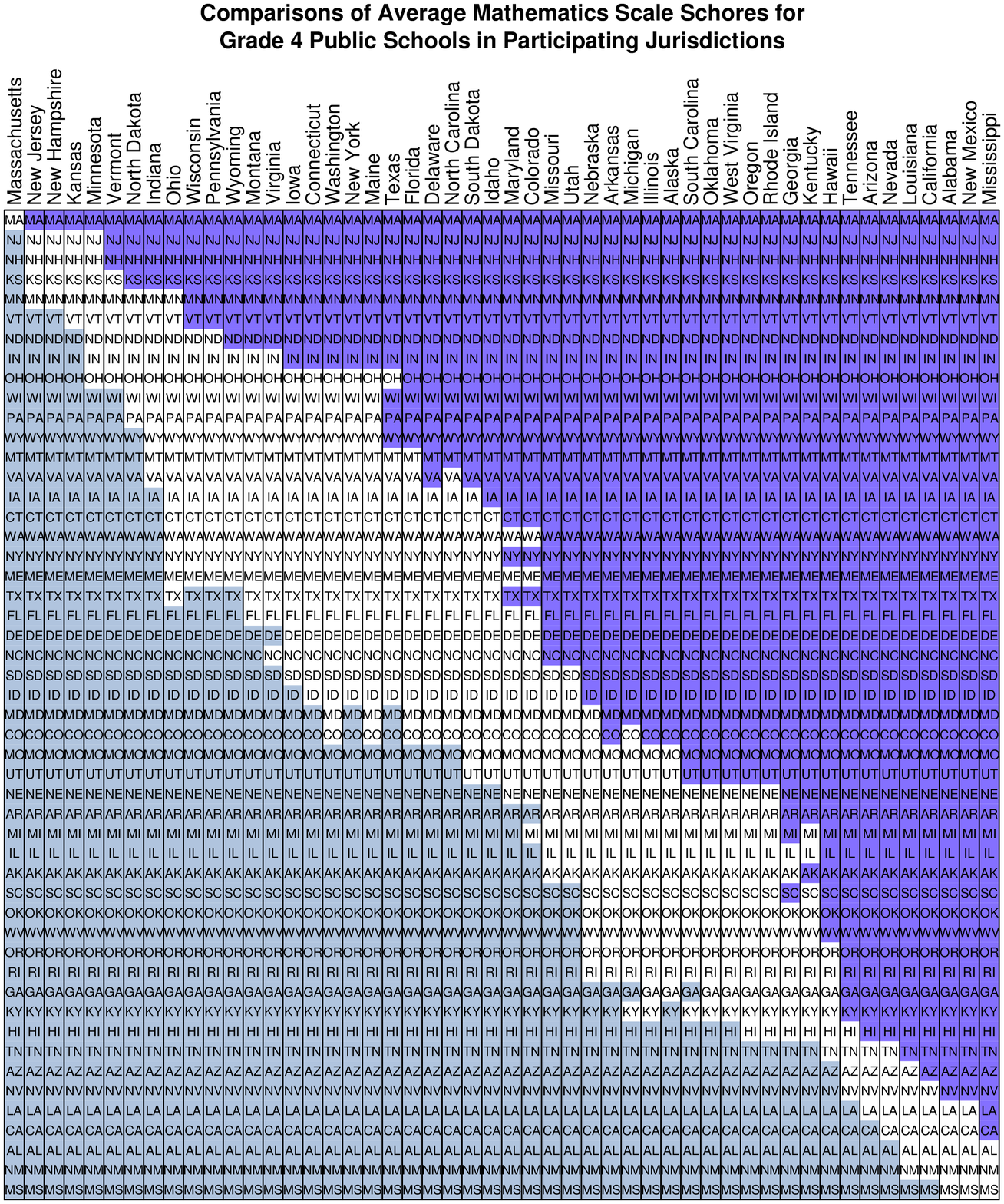}}
  \caption{Posterior inference of NAEP fourth grade mathematics test scores from 2007 by state, fit using a multilevel model.  Compared to the classical multiple comparisons summaries in Figure \ref{naep}, we have fewer cases of ambiguity, that is, more claims with confidence.  This makes sense because the hierarchical model adapts to the setting, which in this case is a high variance ratio and thus little need to worry about noise in the comparisons.}
  \label{naepPosterior}
\end{figure}

One advantage of the Bayesian paradigm
in which models are fit using simulation is that the output is
easy to manipulate in order to examine whatever functions of the
parameters are of interest.  In this case, based on the fitted
multilevel model, we simulate $1000$ draws
of state effect parameters to construct a posterior interval for the
difference in true means for each pair of states.  For the purpose of
comparing to the classical approach, we set a 0.05 cutoff: for each
pair of states $j,k$, we check whether 95\% or more of the simulations
show $\alpha_j>\alpha_k$.  If so---that is, if $\alpha_j>\alpha_k$ for
at least 950 of the 1000 simulations---we can claim with 95\%
confidence that state $j$ outperforms state $k$.  We plot the results
in Figure \ref{naepPosterior}.  States that have effects that are
significantly lower are shaded with light blue, the ones which are
higher are shaded with darker blue, and ones that are not
significantly different are left as white.

Compared to the classical multiple comparisons summaries in Figure
\ref{naep}, the multilevel estimates in Figure \ref{naepPosterior} are
more informative, with more claims with confidence and fewer cells in
the central region where comparisons are not statistically
significant.  The classical procedure overcorrects for multiple
comparisons in this setting where the true differences between states
are large.  (At the other extreme, classical procedures undercorrect
for multiple comparisons when true differences are small, as we
discuss in Section \ref{8schools}.)

When there is evidence for
a multiple comparisons problem, our procedure makes corrections.  When
there is no evidence for a multiple comparisons problem our procedure
is similar to the direct inference without a multiple comparisons
correction.

\subsection{SAT coaching in 8 schools}\label{8schools}

\begin{figure}
  \begin{center}
\small{
    \begin{tabular}{ccccc}
      &\multicolumn{1}{c}{Raw estimate}&\multicolumn{1}{c}{Standard error}& \multicolumn{1}{c}{Bayes}  & \multicolumn{1}{c}{Bayes} \\
      &\multicolumn{1}{c}{of treatment}&\multicolumn{1}{c}{of raw effect} & \multicolumn{1}{c}{posterior} & \multicolumn{1}{c}{posterior}\\
      School&\multicolumn{1}{c}{effect, $y_j$}&
      \multicolumn{1}{c}{estimate, $\sigma_{y\,j}$} & \multicolumn{1}{c}{mean} & \multicolumn{1}{c}{sd}\\\hline
      A & \ 28    & 15  & 11    & 8\\
      B & \ \,\,8 & 10  & \,\,7 & 6\\
      C & $-3$    & 16  & \,\,6 & 8\\
      D & \ \,\,7 & 11  & \,\,7 & 7\\
      E & $-1$    & \ 9 & \,\,5 & 6\\
      F & \ \,\,1 & 11  & \,\,6 & 7\\
      G & \ 18    & 10  & 10    & 7\\
      H & \ 12    & 18  & \,\,8 & 8
    \end{tabular}
}
  \end{center}
  \caption{First two columns of numbers:  Data from the 8-schools experiment of Rubin (1981).  A separate randomized experiment was conducted in each school, and regression analysis gave separate treatment effect estimates (labeled as $y_j$ above) and standard errors (labeled as $\sigma_{y\,j}$).  Effects are on the scale of points in the SAT-Verbal test (which was scored from 200 to 800).  An effect of 8 points corresponds to approximately one additional test item correct.\newline
  Last two columns:  Posterior mean and standard deviation of treatment effects, as estimated from a Bayesian multilevel model.  The evidence is that the effects vary little between schools, hence the estimates are pooled strongly toward the common mean.  None of the comparisons from the Bayesian inference are even close to statistically significant.}
  \label{schools}
\end{figure}

Rubin (1981) discusses an example (reprised in chapter 5 of Gelman et al., 2003) of a meta-analysis of randomized experiments of coaching for the Scholastic Aptitude Test (SAT) in eight high schools in New Jersey.  This example is notable as one of the first fully Bayesian analysis of a hierarchical model and also because there was no evidence in the data of differences between the treatment effects in the different schools.  (And, in fact, the total estimated effects are small.)
%; see Hansen, 2004, for more on estimates of the effects of test coaching programs.)
The first two columns of numbers in Figure \ref{schools} give the
data.  Just to get a sense of the variation, the standard deviation of
the eight school estimates is 10, which is of the same order as the
standard errors.

\paragraph{Classical and Bayesian analysis.}
This is the sort of situation where one might worry about multiple
comparisons.  (In the actual data in Figure \ref{schools}, none of the raw
comparisons happen to be statistically significant, but as we discuss below,
they could be in a replication of the study.)

The hierarchical Bayesian analysis of Rubin (1981) has no multiple
comparisons problems, however.  The group-level variance is estimated
to be low---the marginal maximum likelihood or posterior mode estimate
is zero, and the Bayesian analysis averages over the posterior
distribution, which is largely below 10---and as a result the Bayes
estimates are pooled strongly toward the common mean.

% better prior for the 8 schools??

\paragraph{Simulation study with small effects.}
To get further insight into this example, we perform repeated
simulations of a world in which the true treatment effects in
different schools come from a normal distribution with standard
deviation 5 (a plausible estimate given the data in Figure
\ref{schools}).  For each replication, we simulate eight true values
$\theta_1,\dots,\theta_8$ from this distribution, then simulate data
$y_1,\dots,y_8$ from the eight separate normal distributions
corresponding to each $\theta_j$.  The standard deviations $\sigma_j$
for each of these distributions is given by Figure \ref{schools}.
Relative to the within-group standard deviations, the between-group standard deviation of 5 is small.  We then performed both
classical and hierarchical Bayesian analyses.  For each analysis, we
computed all $(8 \cdot 7)/2=28$ comparisons and count the number that are
statistically significant (that is, where the difference between the
estimates for two schools is more than 1.96 times the standard error
for the difference), and of these, we count the number that have the
correct sign.

We performed 1000 simulations.  Out of these simulations, 7\% of the
classical intervals were statistically significant and, of these, only 63\%
got the sign of the comparison correct.  Multiple comparisons
corrections are clearly necessary here if we want to avoid making
unreliable statements.  By comparison, only 0.5\% of the Bayesian
intervals are statistically significant (with 89\% getting the sign of
the comparison correct).  The shrinkage of the Bayesian analysis has
already essentially done a multiple comparisons correction.

To look at it another way: the classical estimates found at least one
statistically significant comparison in 47\% of our 1000 simulations.
In the Bayesian estimates, this occurred only 5\% of the time.
The Bayesian analysis here uses a uniform prior distribution on the
hyperparameters---the mean and standard deviation of the school
effects---and so it uses no more information than the classical
analysis.  As with a classical multiple comparisons procedure, the
Bayesian inference recognizes the uncertainty in inferences and
correspondingly reduces the number of statistically significant
comparisons.

\paragraph{Simulation study with large effects.}
To get a sense of what happens when effects are more clearly
distinguishable, we repeat the above simulation but assume the true
treatment effects come from a distribution with standard deviation 10.
This time, 12\% of the classical comparisons are statistically
significant, with 86\% of these having the correct sign.  From the
Bayesian analysis, 3\% of the comparisons are statistically
significant, with 96\% of these having the correct sign.  Whether
using classical multiple comparisons or Bayesian hierarchical
modeling, the price to pay for more reliable comparisons is to claim
confidence in fewer of them.

\subsection{Teacher and school effects in NYC schools}
Rockoff (2004) and Kane, Rockoff, and Staiger (2007) analyzed a huge
panel dataset of teachers and children from the New York City school
system to assess the importance of factors such as educational
background, training, and experience in determining the effectiveness
of teachers.  One of the findings was that variation in teacher
``effects'' (we are not interpreting these finding causally) on
student grades was moderately large, about 0.15 standard deviations on
a scale in which standard deviation was calculated using test scores
for all students of a given grade level in the system.  The
researchers, using an approach that approximates the fit from a multilevel
model, learned from the scale of unexplained variation in teacher
effects over time---the residual group-level variance---that teacher
effects are important and often persistent and are largely not
explained by background variables, except for a small improvement in
performance during the first decade of a teacher's career.

More broadly, there has been an increasing push by certain school
districts and policy advocates to use data like these to compare the
``effectiveness'' of individual teachers (or schools) to award merit
pay or provide other incentives or sanctions.  (Grudgingly) leaving
the problematic causal issues aside (see Rubin et al., 2004, for a
relevant discussion), we note that a outstanding methodological problem
thus far has been that analyses used to make such comparisons rarely
if ever address the extreme multiple comparisons problems involved.
This study could have been set up as a multiple comparisons problem,
trying to get appropriate $p$-values for comparing thousands of
teachers or for distinguishing individual teacher effects from zero.
However, we know that there are true differences across teachers and
that teachers should not have no effect (an individual teacher's effect could be negative or
positive but will not be precisely zero), and we therefore should primarily be
concerned with Type S and Type M errors.  Therefore a multilevel model
would be a far more appropriate choice for such analyses---and, in fact, this is essentially what the Kane, Rockoff, and Staiger did.

\subsection{Fishing for significance:  Do beautiful parents have more daughters?}
In an analysis of data from 2000 participants in the U.S. adolescent
health study, Kanazawa (2007) found that more attractive people were
more likely to have girls, compared to the general population: 52\% of
the babies born to people rated ``very attractive'' were girls,
compared to 44\% girls born to other participants in the survey.  The
difference was statistically significant, with a $t$-value of 2.43.
However, as discussed by Gelman (2007), this particular
difference---most attractive versus all others---is only one of the
many plausible comparisons that could be made with these data.
Physical attractiveness in the survey used by this paper was measured
on a five-point scale.  The statistically significant comparison was
between category 5 (``very attractive'') vs.\ categories 1--4.  Other
possibilities include comparing categories 4--5 to categories 1--3
(thus comparing ``attractive'' people to others), or comparing 3--5 to
1--2, or comparing 2--5 to 1.  Moreoever these data are from one of 3
potential survey waves.  Therefore there are 20 possible comparisons
(5 natural comparisons by 4 possible time summaries (wave 1, wave 2,
wave 3, or average).  It is not a surprise that one of this set of
comparisons comes up statistically significant.

In this study, classical
multiple comparisons adjustments may not be such a bad idea, because
actual sex ratio differences tend to be very small---typically less
than 1 percentage point---and so the null hypothesis is approximately
true here.  A simple Bonferroni correction based on our count of 20
possible comparisons would change the critical value from 0.05
to 0.0025 in which case the finding above (with $p$-value of 0.015)
would not be statistically significant.

With a properly designed study, however, multiple comparisons
adjustments would not be needed here either.  To start with, a simple
analysis (for example, linear regression of proportion of girl births
on the numerical attractiveness measure) should work fine.  The
predictor here is itself measured noisily (and is not even clearly
defined) so it would probably be asking too much to look for any more
finely-grained patterns beyond a (possible) overall trend.  More
importantly, the sample size is simply too low here, given what we
know from the literature on sex-ratio differences.  From a classical
perspective, an analysis based on 2000 people is woefully
underpowered and has a high risk of both Type S and Type M errors.

Alternatively, a Bayesian analysis with a reasonably uninformative
prior distribution (with heavy tails to give higher probability to
the possibility of a larger effect) reveals the lack of information
in the data (Gelman and Weakliem, 2009).  In this
analysis the probability that the effect is positive is only 58\%, and the estimated effect size is well under 1 percentage point.
%reveals the data to be much less informative than a reasonable prior distribution

\subsection{Examining impacts across subgroups}
We build on our Infant Health and Development Program example from
Section \ref{illustrative} to illustrate how a multi-site analysis
could be expanded to accommodate subgroup effects as well.  The most
important moderator in the IHDP study was the birth-weight group
designation.  In fact there was reason to believe that children in the
lighter low-birth-weight (less than 2 kg) group might respond differently
to the intervention than children in the heavier (more than 2 kg)
low-birth-weight group.

We expand our
model to additionally allow for differences in treatment effects across
birth weight group,
\begin{equation*}
y_i \sim \Nor(\gamma_{j[i]} + \delta^{\text{L}}_{j[i]}P_i(1-B_i) + \delta^{\text{H}}_{j[i]}P_iB_i,\,\sigma^2_y),
\end{equation*}
Here the treatment effect corresponding
to person $i$'s site (indexed by $j$) depends on whether
the child belongs to the lower low-birth-weight group $\delta^{\text{L}}_{j[i]}$
or the higher low-birth-weight group $\delta^{\text{H}}_{j[i]}$.
This time each of these sets of parameters gets its own distribution
\begin{eqnarray*}
\delta^{\text{L}}_j &\sim& \Nor(\mu_{\text{L}},\sigma_{\delta\,\text{L}}^2)\text{, and}\\
\delta^{\text{H}}_j &\sim& \Nor(\mu_{\text{H}},\sigma_{\delta\,\text{H}}^2).
\end{eqnarray*}
In this case we allow the treatment effects for the lower and higher
low-birth-weight children to have a correlation $\rho$.  Again we have allowed the
intercept, $\gamma_j$, to vary across sites $j$ and have specified prior
distributions for the hyperparameters that should have little to no
impact on our inferences.

\begin{figure}
  \centerline{\psfig{figure=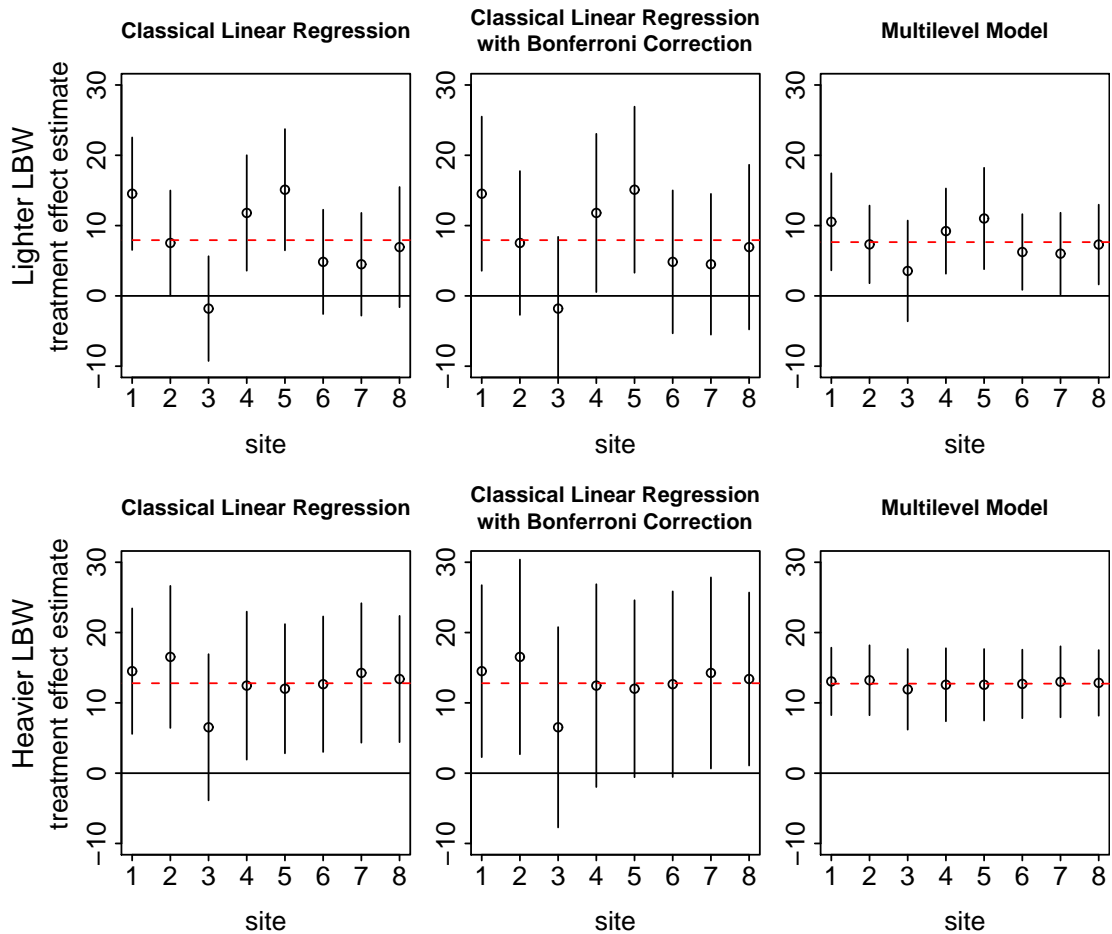,height=5in}}
%  \centerline{\psfig{figure=plot2.eps,width=6.5in}}
  \caption{Treatment effect point estimates and 95\% intervals across the eight IHDP sites now broken down by birth-weight group as well.  The left panel display classical estimates from a linear regression.  The middle panel displays the same point estimates as in the left panel but the confidence intervals have been adjusted to account for a Bonferroni correction.  The right panel displays 95\% intervals and means from the posterior distributions for each of the eight site-specific treatment effects generated by fitting a multilevel model.
}
  \label{ihdp.bybw}
\end{figure}

Figure \ref{ihdp.bybw} plots some results from this model.  The
estimates for the lighter low-birth-weight group are quite volatile in the
classical setting, where the information we have about the
relationship between the sites is ignored.  The Bonferroni correction
serves only to reinforce our uncertainty about these estimates.  On
the other hand, the results from the Bayesian multilevel model for
this group have been shrunk towards the main effect across groups and
thus are less subject to the idiosyncracies that can arise in small
samples.  (The sample sizes across these sites for this group range from
67 to 93.)  The point estimates for the heavier low-birth-weight children are
more stable for all analyses relative to those for the lighter
low-birth-weight group, reflecting that there is generally less
treatment heterogeneity for this group of children.  The classical and
corrected standard errors are larger than for the lighter low-birth-weight
children most likely because the sample sizes across sites for this group
are slighlty smaller (ranging from 35 to 51).

Overall, the results from the Bayesian multilevel analysis are the most 
stable and they lead to substantively different conclusions than the
classical analyses.  None of the Bayesian 95\% intervals even comes close
to covering zero.  This contrasts sharply with the results from the Bonferroni-adjusted classical 
intervals, all of which are quite wide (typically at least twice the width of the Bayesian intervals) and four of which
actually include zero (the other four end quite close to zero).

\section{Multiple outcomes and other challenges}
Similar issues arise when researchers attempt to evaluate the impact
of a program on many different outcomes.  If you look at enough
outcomes, eventually one of them will appear to demonstrate a positive
and significant impact, just by chance.  In theory, multilevel models
can be extended to accommodate multiple outcomes as well.  However
this often requires a bigger methodological and conceptual jump.  The
reason that multilevel models were such a natural fit in the examples
described above is because all the estimated groups effects were
estimates of the same phenomenon.  Thus it was reasonable to assume
that they were \emph{exchangeable}.  Basically this means that we
could think of these estimates as random draws from the same
distribution without any a priori knowledge that one should be bigger
than another. If we had such information then this should be included in the model, and in more complicated settings it is not trivial to set up such a model.

On the other hand, there are some situations when modeling multiple
outcomes simultaneously within a simple multilevel model might be
fairly natural.  For instance, sometimes researchers acquire several
different measures of the same phenomenon (such as educational
achievement, behavioral problems, or attitudes towards related
issues).  It is also common to measure the same attribute at several
different time points over the course of the study.  This might
require a slightly more complicated modeling strategy to account for
trends over time, but is otherwise a reasonable choice.  If many
outcomes have been measured across several disparate domains, however,
more effort may be needed to set up a suitable multilevel model.

%More generally, if you are comparing a set of exchangeable items, such
%as 50 states or 8 schools or 4 machine learning algorithms, it's
%pretty easy to set up a multilevel model.  Some choices need to be
%made---normal distribution or $t$ distribution, mixture model or not,
%equal or unequal variances, etc.---but in practice it's not hard to
%come up with a reasonable model based on one's general understanding
%of the situation.  Even if the items differ in recognizable ways, we
%can still make them exchangeable by including covariates, for example
%statewide poverty levels or previous test scores in the NAEP example.

We illustrate a simple example of multiple outcomes by returning to the IHDP data.
This time we allow treatment effects to vary by site and type of test.
We include eight different types of cognitive tests that were
administered either at year 3 (PPVT-R, Stanford Binet), year 5
(PPVT-R, Weschler Preschool and Primary Scale of Intelligance Revised
verbal and performance subscales), or year 7 (PPVT-R, Weschler
Intelligence Scale for Children verbal and perfomance subscales).
In this new formulation test-specific individual level outcomes
(indexed by $i$) are allowed both site-specific $\gamma^{\text{site}}_j$
and test-specific $\gamma^{\text{test}}_k$ contributions and the
treatment effects, $\delta_l$, are also allowed to vary by site and 
outcome.  Here $l$ indexes site $\times$ test (that is, $j \times k$) 
combinations.
\begin{equation*}
y_i \sim \Nor(\mu + \gamma^{\text{site}}_{j[i]} + \gamma^{\text{test}}_{k[i]} + \delta_{l[i]}P_i,\,\sigma^2_y),
\end{equation*} 
As with the previous models, the site and test-specific intercepts,
$\gamma^{\text{site}}_j$ and $\gamma^{\text{test}}_k$, respectively,
are assumed to follow normal distributions each with mean zero (since the model already includes a
parameter for overall mean, $\mu$) and
each with its own variance.
%, $\sigma_gs$ and $\sigma_gt$, respectively.

What's more interesting here is the model for the $\delta_l$,
\begin{equation*}
\delta_l \sim \Nor(\delta_0 + \delta^{\text{site}}_{j[l]} + \delta^{\text{test}}_{k[l]},\,\sigma^2_{\delta}).
\end{equation*}
Here the site-specific contributions to the treatment effects
$\delta^{\text{site}}_l[j]$ are simply modeled with a normal
distribution with mean zero and separate variance component.  However the
test-specific contributions, $\delta^{\text{test}}_k$, are allowed to
systematically vary based on the age at which the test was taken and
whether or not the test measures verbal skills
\begin{equation*}
\delta^{\text{test}}_k \sim \Nor(\phi_1 \text{year5} + \phi_2 \text{year7} + \phi_3 \text{verbal},\, \sigma^2_{\delta_{\text{test}}}).
\end{equation*}
This last piece of the model increases the plausibility of the
exchangeability assumption for the test scores.  We also simplify
the model by first standardizing each of the test scores (to have mean
0 and standard deviation 1) within the sample.

\begin{figure}
%  \centerline{\includegraphics[scale=.65]{figure=site.by.test.eps}}
  \centerline{\psfig{figure=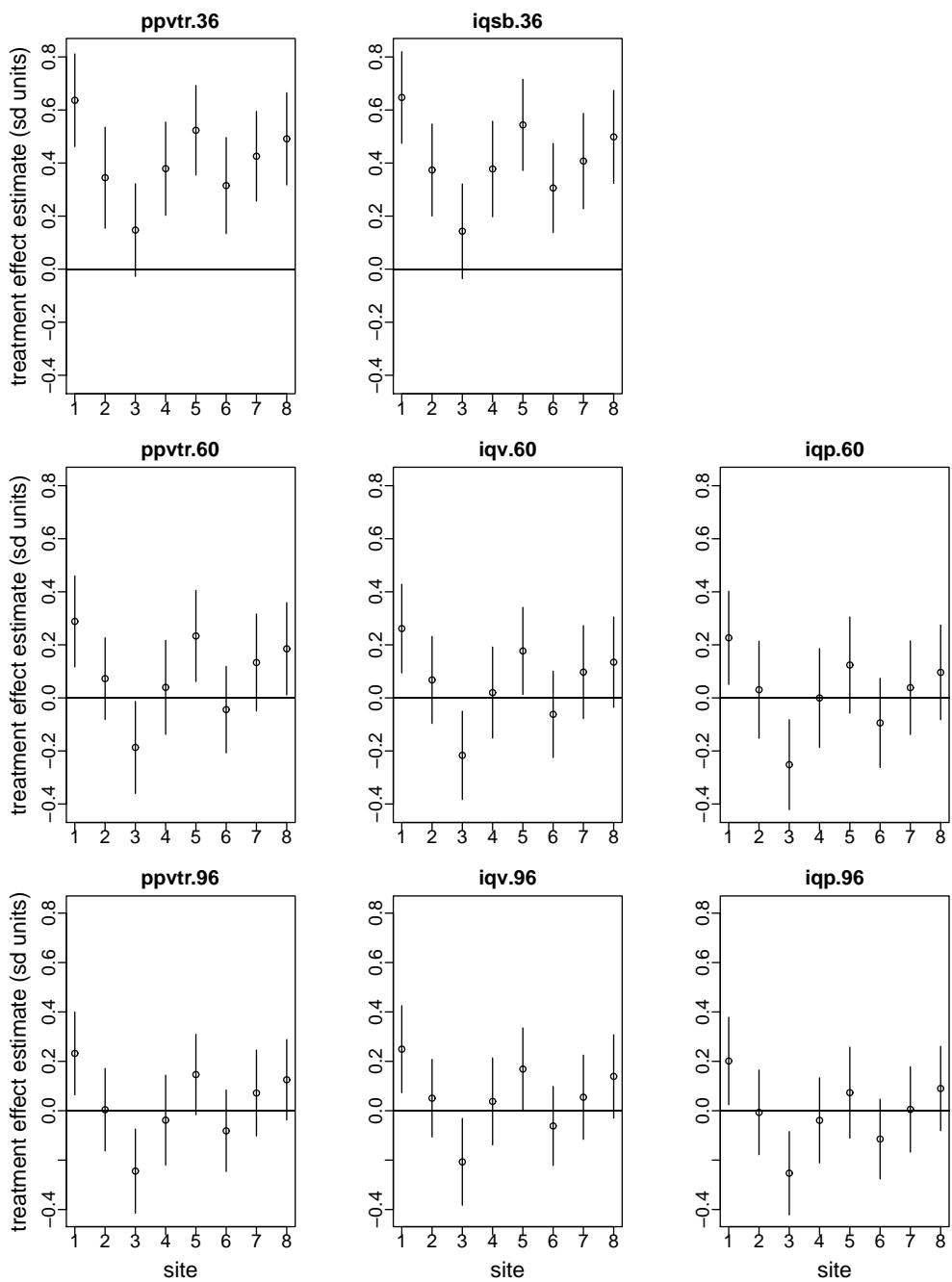, height=7in}}
  \caption{Example of inference for multiple outcomes.  Estimated 95\% intervals and medians from the posterior distributions for each of 64 site-and-test-specific treatment effects in the Infant Health
and Development Program analysis, from by fitting a multilevel model.
}
  \label{ihdp.by.site.and.test}
\end{figure}

The results from this model are displayed in Figure
\ref{ihdp.by.site.and.test}.  Each row of the figure corresponds to the 
test score outcome from a different year (row 1 for age 3, row 2 for age 5,
row 3 for age 8).  Within each plot, 95\% intervals are displayed for
each site.  The treatment effects are larger on average for tests taken
directly as the intervention ended at age 3.  Similar patterns appear across 
sites for each outcome---this phenomenon reflects an assumption that could 
be relaxed by including site by outcome interactions.

\begin{figure}
  \centerline{\psfig{figure=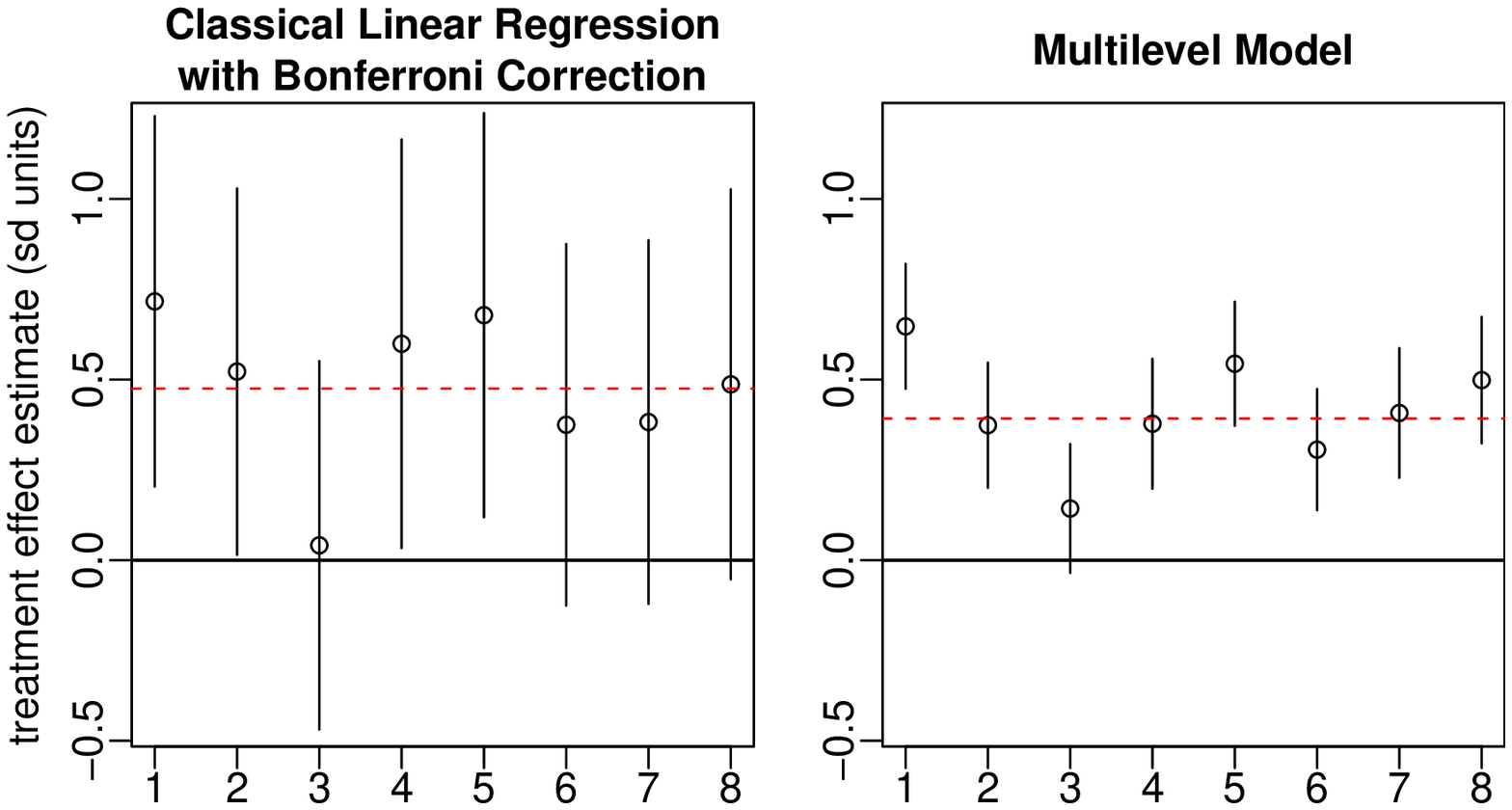, height=2.5in, width=5in}}
  \caption{Comparison of treatment effects by site between classical linear regression with Bonferroni correction and multilevel model fit for the multiple-outcomes analysis of the Infant Health
and Development Program (see Figure \ref{ihdp.by.site.and.test}).  Both account for eight sites and eight outcomes though only one outcome, Stanford Binet IQ score at age 36 months, is displayed.
        \label{bonf.vs.mlm}}
\end{figure}

A comparison to a classical correction such as a Bonferroni adjustment
is even more extreme in this setting, as illustrated in Figure \ref{bonf.vs.mlm}.
When taking all eight outcomes into consideration, the Bonferroni 
correction applied to the classic linear regression fit results in even 
more extreme uncertainty bounds than in our original example because now
there are 64 comparisons rather than simply eight.  The multilevel model
estimates are now additionally shrunk towards the grand mean across outcomes,
after adjusting for differences in mean test scores attributable to the
year they were administered and the type of test.  This shrinkage causes 
the estimates to be more conservative than they previously were.  However
the overall precision of the multilevel model results is vastly superior 
to the Bonferroni-adjusted intervals.

%SHOULD WE BE INSTEAD COMPARING AT LEAST SOME OF THESE PLOTS WITH WHAT
%WOULD HAVE HAPPENED WITH SEPARATE ANALYSES FOR EACH OUTCOME AND PERHAPS
%BONFERRONI CORRECTIONS?
%** OK

\paragraph{Further complications}

%Also, at a technical level, when the number of groups is smaller, multilevel inferences become more sensitive to the prior distribution on the variance parameter (Gelman, 2006).

Harder problems arise when modeling multiple comparisons that have
more structure.  For example, suppose we have 5 outcome measures, 3 varieties of treatments, and subgroups
classified by 2 sexes and 4 racial groups.  We would not want to model
this $2\times 3 \times 4 \times 5$ structure as 120 exchangeable groups.  Even in these more complex situations, we think multilevel modeling
should and will eventually take the place of classical multiple
comparisons procedures.  After all, classical
multiple comparisons themselves assumes
exchangeability in the sense of treating all the different comparisons
symmetrically.  And so, in either case, further work is neeeded for the method to match the problem structure.  For large problems, there can be more data for estimating variance parameters in multilevel models (this is sometimes called the blessing of dimensionality).  Similarly, classical procedures may have the potential to adaptively vary tuning parameters in large, complex structures.

\section{Conclusion}
Multiple comparisons can indeed create problems and it is useful to
address the issue.  However, statistical methods must be mapped to the
applied setting.  Classical Type 1 and Type 2 errors and false
discovery rates are based on the idea that the true effect could
really be zero (see Johnstone and Silverman, 2004, and Efron, 2006,
for connections between these ideas and hierarchical Bayes methods).
Effects that are truly zero (not just ``small'') can make sense in
genetics (Efron and Tibshirani, 2002) but are less likely in social
science or education research.  We prefer to frame the issue in terms
of Type S or Type M errors.

Therefore, when doing social science or program evaluation, we do not
recommend classical methods that alter $p$-values or (equivalently)
make confidence intervals wider.  Instead, we prefer multilevel
modeling, which shifts point estimates and their corresponding
intervals closer to each other (that is, performs partial pooling)
where necessary---especially when much of the variation in the data
can be explained by noise.  Therefore fitting the multilevel model
should result in the positive externality of yielding more reliable
estimates for individual groups.

We recognize that multilevel modeling can prove to be more of a
challenge for complicated structures.
More research needs to be done in this area.  However we believe it is
more worthwhile to invest research time and effort towards expanding
the multilevel model framework than to invest in classical \mc
adjustments that start from a perspective on the problem to which we
do not adhere.

%In this way few claims can be made with confidence.   -- this not
% always true as in our ihdp example above.  fewer claims about the
% differences between sites could be made but actually more claims
% could be made after the partial pooling about site effects being
% statistically significantly different from 0
%By comparison, classical intervals can have Type S error rates as high
%as 50\%: then, multiple comparisons can pose a real threat.

Applied researchers may balk at having to learn to fit a different
kind of model.  However, functions for fitting multilevel models are
now available in many statistical software packages; therefore,
implementing our suggestions should not be overly burdensome.
Moreover, multiple comparisons problems arise frequently in
research studies in which participants have been clustered because of
interest in examining differences across these program sites, schools,
cities, etc; arguably, data from these types of studies should 
be fit using a multilevel model anyway to correctly reflect the
within-group correlation structure of the errors.
Thus the multilevel model will not only yield better results than
the simplest multiple comparisons corrections, it should not pose a
greater burden than performing one of the fancier types of classical
types of multiple comparisons corrections. 

%For example, analyzing 16
%schools or 16 outcomes is simple enough, but a $2^4$ structure (e.g.,
%2 groups $\times$ 2 outcomes $\times$ 2 flavors of treatment $\times$
%2 time points) is more of a research problem.

%The key concern about multiple comparisons is when the noise level is
%high, so that classical Type S error rates will be high.  Having {\em
%many} comparisons (100, or 1000, or whatever) is not a problem at all
%in the context of multilevel modeling.  This gets back to our
%willingness to be wrong occasionally.  If we are testing 1000
%comparisons, and there's really nothing much going on, then the
%multilevel analysis will do lots of pooling, and the result is to make
%very few claims with confidence.

\section*{References}

\noindent

\bibitem Almond, R. G., Lewis, C., Tukey, J. W., and Yan, D. (2000).  Displays for comparing a given state to many others.  {\em American Statistician} {\bf 34}, 89--93.

\bibitem Benjamini, Y., and Hochberg, Y. (1995).  Controlling the false discovery rate:  a practical and
powerful approach to multiple testing.  {\em Journal of the Royal Statistical Society B} {\bf 57}, 289--300.

\bibitem Benjamini, Y., and Yekutieli, D. (2001).  The control of the false discovery rate in multiple testing under dependency.  {\em Annals of Statistics} {\bf 29}, 1165--1188.

%\bibitem Blackman, C. F., Benane, S. G., Elliott, D. J., House, D. E., and Pollock, M. M. (1988). Influence of electromagnetic fields on the efflux of calcium ions from brain tissue in vitro:  a three-model analysis consistent with the frequency response up to 510 Hz. {\em Bioelectromagnetics} {\bf 9}, 215--227.

%\bibitem Bowen, J. (1999).  Faith healing: can prayer do anything more than make you feel better?  {\it Salon}, November 3, {\tt www.salon.com/health/feature/1999/11/03/prayer}

%\bibitem Demsar, J. (2006).  Statistical comparisons of classifiers over multiple data sets.  {\em Journal of Machine Learning Research} {\bf 7}, 1--30.

\bibitem Efron, B. (2006).  Size, power, and false discovery rates.  Technical report, Department of Statistics, Stanford University.

\bibitem Efron, B., and Morris, C. (1975).  Data analysis using Stein's estimator and its generalizations.  {\em Journal of the American Statistical Association} {\bf 70}, 311--319.

\bibitem Efron, B., and Tibshirani, R. (2002). Empirical Bayes methods and false discovery rates for microarrays.  {\em Genetic Epidemiology} {\bf 23}, 70--86.

%\bibitem Gelman, A. (2006).  Prior distributions for variance
%  parameters in hierarchical models.  {\em Bayesian Analysis} {\bf 1},
%  514--534.

\bibitem Gelman, A. (2006).  Prior distributions for variance parameters in hierarchical models. {\em Bayesian Analysis} {\bf 1}, 515--533.

\bibitem Gelman, A. (2007).  Letter to the editor regarding some papers of Dr.\ Satoshi Kanazawa.  {\em Journal of Theoretical Biology} {\bf 245}, 597--599.

%\bibitem Gelman, A. (2007b).  Struggles with survey weighting and regression modeling (with discussion).  {\em Statistical Science}.

\bibitem Gelman, A., Carlin, J. B., Stern, H. S., and Rubin, D. B. (2003).
{\em Bayesian Data Analysis}, second edition.  London:  CRC Press.

\bibitem Gelman, A., and Hill, J. (2007).  {\em Data Analysis Using Regression and Multilevel/Hierarchical Models}.  Cambridge University Press.

%\bibitem Gelman, A., and Jakulin, A. (2007).  Bayes:  liberal, radical, or conservative?  {\em Statistica Sinica}.

%\bibitem Gelman, A., and Nolan, D. (2002).  {\em Teaching Statistics:  A Bag of Tricks}.  Oxford University Press.

%\bibitem Gelman, A., and Park, D. K. (2007).  Splitting a predictor at the upper quarter or third and the lower quarter or third.  Technical report, Department of Statistics, Columbia University.

\bibitem Gelman, A., and Stern, H. S. (2006).  The difference between ``significant'' and ``not significant'' is not itself statistically significant.  {\em American Statistician} {\bf 60}, 328--331.

\bibitem Gelman, A., and Tuerlinckx, F. (2000).  Type S error rates for classical and Bayesian single and
multiple comparison procedures.  {\em Computational Statistics} {\bf 15},
373--390.

\bibitem Gelman, A., and Weakliem, D. (2009).  Of beauty, sex, and power:
  statistical challenges in estimating small effects.  {\em American Scientist} {\bf 97}, 310--316.

\bibitem Genovese, C., and Wasserman, L. (2002). Operating characteristics and extensions of the false discovery rate procedure.  {\em Journal of the Royal Statistical Society: Series B (Statistical Methodology} {\bf 64}, 499--517.

\bibitem Grant, G. R., Liu, J., and Stoeckert, C. J. (2005). A practical false discovery rate approach to identifying patterns of differential experession in microarray data.  {\em Bioinformatics} {\bf 21}, 2684--2690.

\bibitem Greenland, S., and Robins, J. M. (1991).  Empirical-Bayes adjustments for multiple comparisons are sometimes useful.  {\em Epidemiology} {\bf 2}, 244--251.

\bibitem Hansen, B. B. (2004). Full matching in an observational study of coaching for the SAT. {\em Journal of the American Statistical Association} {\bf 99}, 609--619.

%\bibitem Harris, W. S., Gowda, M., Kolb, J. W., Strychacz, C. P., Vacek, J. L., Jones, P. G., Forker, A., O'Keefe, J. H., and McCallister, B. D. (1999).  A randomized, controlled trial of the effects of remote, intercessory prayer on outcomes in patients admitted to the coronary care unit.  {\em Archives of Internal Medicine} {\bf 159}, 2273--2278.

\bibitem Hsu, J. C. (1996). {\em Multiple Comparisons: Theory and Methods}. London: Chapman and Hall.

\bibitem Infant Health and Development Program (1990).  Enhancing the outcomes of low-birth-weight, premature infants.  A multisite, randomized trial.  The Infant Health and Development Program.  {\em Journal of the American Medical Association} {\bf 263}, 3035--3042.

\bibitem James, W., and Stein, C. (1960).  Estimation with quadratic loss.  In {\em Proceedings of the Fourth Berkeley Symposium on Mathematical Statistics and Probability} {\bf 1}, ed. J. Neyman, 361--380.  Berkeley:  University of California Press.

\bibitem Johnstone, I., and Silverman, B. (2004).  Needles and straw in a haystacks: empirical Bayes approaches
to thresholding a possibly sparse sequence.  {\em Annals of Statistics} {\bf 32}, 1594--1649.

\bibitem Kanazawa, S. (2007).  Beautiful parents have more daughters:  a further implication of the generalized Trivers-Willard hypothesis.  {\em Journal of Theoretical Biology}.

\bibitem Kane, T. J., Rockoff, J. E., and Staiger, D. O. (2007).  What does certification tell us about teacher effectiveness? Evidence from New York City.  {\em Economics of Education Review}.

\bibitem Krantz, D. H. (1999).  The null hypothesis testing controversy in psychology.  {\em Journal of the American Statistical Association} {\bf 94},  1372--1381.

\bibitem Louis, T. A. (1984). Estimating a population of parameter values using Bayes and empirical Bayes methods.  {\em Journal of the American Statistical Association} {\bf 79}, 393--398.

\bibitem National Center for Education Statistics (1997).
1996 NAEP comparisons of average scores for participating jurisdictions.  Washington, D.C.:  Government Printing Office.

\bibitem Poole, C. (1991).  Multiple comparisons? No problem!  {\em Epidemiology} {\bf 2}, 241--243.

\bibitem Rockoff, J. (2004).  The impact of individual teachers on student achievement: evidence from panel data.  {\em American Economic Review}, Papers and Proceedings, May.

\bibitem Rubin, D. B. (1981).  Estimation in parallel randomized experiments.
{\em Journal of Educational Statistics} {\bf 6}, 377--401.

\bibitem Rubin, D. B., Stuart, E. A., and Zanutto, E. L. (2004).  A potential outcomes 
view of value-added assessment in education.  {\em Journal of Educational and Behavioral 
Statistics} {\bf 29}, 103--116.

\bibitem Spiegelhalter, D., Thomas, A., Best, N., Gilks, W., and Lunn, D.
(1994, 2002).
BUGS:  Bayesian inference using Gibbs sampling.  MRC Biostatistics Unit, Cambridge, England.

\bibitem Tukey, J. W. (1953).  The problem of multiple comparisons.  {\em Mimeographed notes}, Princeton University.

\bibitem Wainer, H. (1986).  Five pitfalls encountered when trying to compare states on their SAT scores.  {\em Journal of Educational Statistics} {\bf 11}, 239--244.

\bibitem Wainer, H. (1996). Using trilinear plots for NAEP state data.
{\em Journal of Educational Measurement} {\bf 33}, 41--55.

\bibitem Wainer, H., Hambleton, R. K., and Meara, K. (1999). Alternative displays for communicating NAEP results: a redesign and validity study. {\em Journal of Educational Measurement} {\bf 36}, 301--335.

\bibitem Westfall, P. H. and Young, S. S. (1993). {\em Resampling-Based Multiple Testing: Examples and Methods for P-Value Adjustment}.  New York:  Wiley.

\end{document}